\documentclass[aps,prd,twocolumn,superscriptaddress]{revtex4}
\usepackage{epsfig,epsf}
\usepackage{amsmath}
\usepackage{amsthm}
\usepackage{amsfonts}
\usepackage{amssymb}
\usepackage{dsfont}
\usepackage{multirow}
\usepackage{appendix}
\usepackage{slashed}
\usepackage[active]{srcltx}
\usepackage{psfrag}

\begin{document}

\title{Strong decays of double-charmed pseudoscalar and scalar $cc\overline{u%
}\overline{d}$ tetraquarks}
\date{\today}
\author{S.~S.~Agaev}
\affiliation{Institute for Physical Problems, Baku State University, Az--1148 Baku,
Azerbaijan}
\author{K.~Azizi}
\affiliation{Department of Physics, Do\v{g}u\c{s} University, Acibadem-Kadik\"{o}y, 34722
Istanbul, Turkey}
\affiliation{Department of Physics, University of Tehran, North Karegar Ave., Tehran 14395-547, Iran}
\author{H.~Sundu}
\affiliation{Department of Physics, Kocaeli University, 41380 Izmit, Turkey}

\begin{abstract}
The strong decays of the pseudoscalar and scalar double-charmed tetraquarks $%
T_{cc;\overline{u}\overline{d}}^{+}$ and $\widetilde{T}_{cc;\overline{u}%
\overline{d}}^{+}$ are investigated in the framework of the QCD sum rule
method. The mass and coupling of these exotic four-quark mesons are
calculated in the framework of the QCD two-point sum rule approach by taking
into account vacuum condensates of the quark, gluon, and mixed local
operators up to dimension 10. Our results for masses $m_{T}=(4130~\pm 170)~%
\mathrm{MeV} $ and $m_{\widetilde{T}}=(3845~\pm 175)~\mathrm{MeV}$
demonstrate that these tetraquarks are strong-interaction unstable
resonances and decay to conventional mesons through the channels $T_{cc;%
\overline{u}\overline{d}}^{+} \to D^{+}D^{\ast }(2007)^{0},~D^{0}D^{\ast
}(2010)^{+}$ and $\widetilde{T}_{cc;\overline{u}\overline{d}}^{+}\to
D^{+}D^{0}$. Key quantities necessary to compute the partial width of these
decay modes, i.e., the strong couplings of two $D$ mesons and a
corresponding tetraquark $g_i,~i=1,2$, and $G$ are extracted from the QCD
three-point sum rules. The full width $\Gamma _{T}=(129.9\pm 23.5)~\mathrm{%
MeV}$ demonstrates that the tetraquark $T_{cc;\overline{u}\overline{d}}^{+}$
is a broad resonance, whereas the scalar exotic meson with $\Gamma _{%
\widetilde{T}}=(12.4\pm 3.1)~\mathrm{MeV}$ can be classified as a relatively
narrow state.
\end{abstract}

\maketitle


\section{Introduction}

\label{sec:Int}
Double-charmed tetraquarks as exotic mesons are already on agenda of
high-energy physics. Their properties were studied in a more general context
of double-heavy mesons built of a heavy diquark $QQ$ and heavy or light
antidiquarks \cite{Ader:1981db,Lipkin:1986dw,Zouzou:1986qh,Carlson:1987hh}.
A main question addressed in these basic papers was whether such 4-quarks
can form bound states or exist as unstable resonances. It was demonstrated
that exotic mesons $QQ\overline{q}\overline{q}$ might be stable provided
that the mass ratio of constituent quarks $m_{Q}/m_{q}$ is large enough. In
this sense, tetraquarks with a diquark $bb$ are more promising candidates to
stable exotic mesons than ones containing a $bc$ or $cc$ pair. In fact, the
isoscalar $J^{P}=1^{+}$ tetraquark $T_{bb;\bar{u}\bar{d}}^{-}$ is expected
to lie below the two $B$-meson threshold and is strong-interaction stable
state \cite{Carlson:1987hh}. The situation with $T_{bc;\bar{q}\bar{q}%
^{\prime }}$ and $T_{cc;\bar{q}\bar{q}^{\prime }}$ is not quite clear; they
may exist as either bound or resonant states.

In the following years the chiral quark model, dynamical and relativistic
quark models, and other theoretical schemes of high-energy physics were used
to calculate spectroscopic parameters of the double-charmed tetraquarks \cite%
{Pepin:1996id,Cui:2006mp,Vijande:2006jf,Ebert:2007rn}. Production of these
particles in ion, proton-proton, and electron-positron collisions, in $B_{c}$
and $\Xi _{bc}$ decays was investigated as well \cite%
{SchaffnerBielich:1998ci,DelFabbro:2004ta,Lee:2007tn,Hyodo:2012pm,Esposito:2013fma}%
. In the framework of the QCD sum rule method the axial-vector tetraquarks $%
QQ\bar{u}\bar{d}$ were explored in Ref.\ \cite{Navarra:2007yw}. In
accordance with obtained results the mass of $T_{bb;\bar{u}\bar{d}}^{-}$ is
below the open bottom threshold and, hence, it cannot decay directly to
conventional mesons. Within the same method tetraquarks with quantum numbers
$J^{P}=0^{-},~0^{+},~1^{-}$ and $1^{+}$, and the quark content $QQ\bar{q}%
\bar{q}$ were studied in Ref.\ \cite{Du:2012wp}.

Recent intensive investigations of double-heavy tetraquarks were inspired by
the discovery of double-charmed baryon $\Xi _{cc}^{++}=ccu$ \cite%
{Aaij:2017ueg}. The mass of this particle was utilized as input information
in a phenomenological model to evaluate masses of the tetraquarks $T_{bb;%
\overline{u}\overline{d}}^{-}$ and $T_{cc;\overline{u}\overline{d}}^{+}$
\cite{Karliner:2017qjm}. It was confirmed once more that the axial-vector
isoscalar state $T_{bb;\overline{u}\overline{d}}^{-}$ is stable against
strong and electromagnetic interactions, whereas the tetraquark $T_{cc;%
\overline{u}\overline{d}}^{+}$ can decay to $D^{0}D^{\ast +}$ mesons. A
conclusion on a stable nature of $T_{bb;\overline{u}\overline{d}}^{-}$ was
drawn also in Refs.\ \cite{Eichten:2017ffp,Agaev:2018khe}.

The spectroscopic parameters and widths of the double-charmed pseudoscalar
tetraquarks $T_{cc;\overline{s}\overline{s}}^{++}$ and $T_{cc;\overline{d}%
\overline{s}}^{++}$, which bear two units of the electric charge were
calculated in Ref.\ \cite{Agaev:2018vag}. Obtained results showed that these
exotic mesons are rather broad resonances. Various aspects of double-charmed
tetraquarks were analyzed also in the publications \cite%
{Wang:2017dtg,Hyodo:2017hue,Yan:2018gik,Luo:2017eub,Ali:2018ifm}.

In the present work we investigate the pseudoscalar and scalar tetraquarks $%
T_{cc;\overline{u}\overline{d}}^{+}$ and $\widetilde{T}_{cc;\overline{u}%
\overline{d}}^{+}$. First, we calculate their spectroscopic parameters in
the context of the QCD two-point sum rule method by taking into account
nonperturbative contributions up to dimension ten. Our studies demonstrate
that these exotic mesons are unstable resonances, and decay strongly to
conventional mesons. The kinematically allowed decay modes $T_{cc;\overline{u%
}\overline{d}}^{+}\rightarrow D^{+}D^{\ast }(2007)^{+}$, $T_{cc;\overline{u}%
\overline{d}}^{+}\rightarrow D^{0}D^{\ast }(2010)^{+}$, and $\widetilde{T}%
_{cc;\overline{u}\overline{d}}^{+}\rightarrow D^{0}D^{+}$ are analyzed and
their partial widths are found. To this end, we consider the strong
couplings of two $D$ mesons and tetraquarks, which are key quantities of the
analysis, and extract their values from the three-point QCD sum rules.
Obtained predictions are used to estimate the full width of the four-quark
mesons $T_{cc;\overline{u}\overline{d}}^{+}$ and $\widetilde{T}_{cc;%
\overline{u}\overline{d}}^{+}$.

This work has the following structure. In Sec.\ \ref{sec:Masses}, we
calculate the mass and coupling of the tetraquarks $T_{cc;\overline{u}%
\overline{d}}^{+}$ and $\widetilde{T}_{cc;\overline{u}\overline{d}}^{+}$.
Here, we provide details of calculations for the pseudoscalar state $T_{cc;%
\overline{u}\overline{d}}^{+}$ and write down final predictions for $%
\widetilde{T}_{cc;\overline{u}\overline{d}}^{+}$. Section \ref{sec:Decays}
is devoted to analysis of strong decays of the tetraquarks. For these
purposes, we evaluate the couplings $g_{1}(q^{2})$, $g_{2}(q^{2})$ and $%
G(q^{2})$ corresponding to relevant strong vertices, and find the fit
functions to extrapolate sum rule predictions to the relevant $D$ mesons'
mass shell. \ These strong couplings are utilized to evaluate partial width
of decay processes. Our conclusions are presented in Sec.\ \ref{sec:Conc}.


\section{Mass and coupling of the pseudoscalar and scalar tetraquarks $T_{cc;%
\overline{u}\overline{d}}^{+}$ and $\widetilde{T}_{cc;\overline{u}\overline{d%
}}^{+}$}

\label{sec:Masses}
As it has been noted above, the mass and coupling of the tetraquarks $T_{cc;%
\overline{u}\overline{d}}^{+}$ and $\widetilde{T}_{cc;\overline{u}\overline{d%
}}^{+}$ (in what follows denoted by $T$ and $\widetilde{T}$, respectively)
can be evaluated by means of the QCD two-point sum rule method. The
essential component of this approach is the interpolating current, which
should be composed of relevant diquark fields and has the quantum numbers of
the original particle. There are different currents that meet these
requirements \cite{Du:2012wp}. For the pseudoscalar tetraquark $T$ with two
identical $c$ quarks we choose a structure made of the heavy pseudoscalar
and light scalar diquarks
\begin{equation}
J(x)=c_{a}^{T}(x)Cc_{b}(x)\overline{u}_{a}(x)\gamma _{5}C\overline{d}%
_{b}^{T}(x).  \label{eq:Curr1}
\end{equation}%
The current $J(x)$ has the symmetric color structure and belongs to the
sextet representation of the color group. The state $T$ with structure (\ref%
{eq:Curr1}) is a $\overline{u}\overline{d}$ member of the multiplet of
pseudoscalar $cc$ tetraquarks while others are the four-quark mesons $T_{cc;%
\overline{s}\overline{s}}^{++}$ and $T_{cc;\overline{d}\overline{s}}^{++}$.
The present investigation allows us to add the new particle $T$ to the list
of double-charmed pseudoscalar tetraquarks.

The interpolating current for the scalar tetraquark $\widetilde{T}$ can be
constructed from the heavy and light axial-vector diquark fields \cite%
{Wang:2017dtg}%
\begin{equation}
\widetilde{J}(x)=\epsilon \widetilde{\epsilon }[c_{b}^{T}(x)C\gamma _{\mu
}c_{c}(x)][\overline{u}_{d}(x)\gamma ^{\mu }C\overline{d}_{e}^{T}(x)],
\label{eq:Curr2}
\end{equation}%
where $\epsilon \widetilde{\epsilon }=\epsilon ^{abc}\epsilon ^{ade}$. In
expressions above, $a,b,c,d$, and $e$ are color indices, and $C$ is the
charge-conjugation operator.

The QCD two-point sum rules to evaluate the spectroscopic parameters of the
tetraquark $T$ should be derived from the correlation function
\begin{equation}
\Pi (p)=i\int d^{4}xe^{ip\cdot x}\langle 0|\mathcal{T}\{J(x)J^{\dag
}(0)\}|0\rangle .  \label{eq:CF1}
\end{equation}%
After replacement $J(x)\rightarrow \widetilde{J}(x)$ a similar correlator
can be written down for the second particle $\widetilde{T}$ as well. Below
we give details of calculations for the mass $m_{T}$ and coupling $f_{T}$,
and provide only final results for $\widetilde{T}$.

To extract the desired sum rules from the correlation function $\Pi (p)$,
one has, first of all, to express it in terms of the tetraquarks' physical
parameters and, in this way determine their phenomenological side $\Pi ^{%
\mathrm{Phys}}(p)$. The function $\Pi ^{\mathrm{Phys}}(p)$ can be derived by
inserting into the correlation function $\Pi (p)$ a full set of relevant
states, carrying out integration over $x$ in Eq.\ (\ref{eq:CF1}), and
isolating a contribution of the ground-state particle $T$. In this process
we accept an assumption on the dominance of a tetraquark term in the
phenomenological side, which for multiquark hadrons should be applied with
some caution. The reason is that an interpolating current used in such
calculations couples not only to a multiquark hadron, but also to a relevant
two-hadron continuum, which may obstruct the multiquark signal \cite%
{Kondo:2004cr}. But direct subtraction of the two-hadron contributions from
the correlator leads to wrong results and conclusions \cite{Lee:2004xk}. To
solve this problem, the authors in Ref.\ \cite{Lee:2004xk} utilized an
alternative way and computed explicitly a coupling of a two-hadron continuum
with a pentaquark current, and demonstrated that these effects constitute
less than $10\%$ of the sum rules.

A more general method to treat similar contributions in the sum rules
involving tetraquarks was used in Ref. \cite{Wang:2015nwa}. It turns out
that two-meson continuum contributions give rise to the finite width $\Gamma
(p^{2})$ of the tetraquark, which can be taken into account by modifying its
propagator. In the sum rules, this modification leads to rescaling of the
tetraquark's coupling, while the mass remains unchanged. Our calculations
showed that even for the tetraquarks with the full width of a few hundred
mega-electron-volts, the two-meson continuum changed the coupling
approximately by $(5-7)\%$ \cite{Agaev:2018vag,Sundu:2018nxt}. This
uncertainty does not exceed the accuracy of the sum rule calculations
themselves; therefore to derive $\Pi ^{\mathrm{Phys}}(p)$ one can neglect it
and use the zero-width single-pole approximation.

Then for $\Pi ^{\mathrm{Phys}}(p)$ we get
\begin{equation}
\Pi ^{\mathrm{Phys}}(p)=\frac{\langle 0|J|T(p)\rangle \langle
T(p)|J^{\dagger }|0\rangle }{m_{T}^{2}-p^{2}}+\ldots ,  \label{eq:Phys1}
\end{equation}%
which contains the contribution of the ground-state particle written down
explicitly as well as effects due to higher resonances and continuum states:
the latter in Eq.\ (\ref{eq:Phys1}) is denoted by dots.

The correlation function $\Pi ^{\mathrm{Phys}}(p)$ can be recast into a more
simple form if one introduces the matrix element of the pseudoscalar
tetraquark
\begin{equation}
\langle 0|J|T(p)\rangle =\frac{f_{T}m_{T}^{2}}{2m_{c}}.  \label{eq:ME1}
\end{equation}%
Then, we find
\begin{equation}
\Pi ^{\mathrm{Phys}}(p)=\frac{1}{4m_{c}^{2}}\frac{f_{T}^{2}m_{T}^{4}}{%
m_{T}^{2}-p^{2}}+\ldots  \label{eq:Phys1a}
\end{equation}%
In general, to continue calculations one should choose in $\Pi ^{\mathrm{Phys%
}}(p)$ some Lorentz structure and fix the corresponding invariant amplitude.
Because in the case under discussion $\Pi ^{\mathrm{Phys}}(p)$ has the
trivial structure which is proportional to $I$, the amplitude $\Pi ^{\mathrm{Phys}%
}(p^{2})$ equals the function from Eq.\ (\ref{eq:Phys1a}).

The QCD side of the sum rules $\Pi ^{\mathrm{OPE}}(p)$ can be found by
computing the correlation function in terms of the quark propagators. To
this end, we insert the interpolating current $J(x)$ to the expression (\ref%
{eq:CF1}) and after contracting the relevant quark fields find
\begin{eqnarray}
&&\Pi ^{\mathrm{OPE}}(p)=i\int d^{4}xe^{ipx}\mathrm{Tr}\left[ \gamma _{5}%
\widetilde{S}_{d}^{b^{\prime }b}(-x)\gamma _{5}S_{u}^{a^{\prime }a}(-x)%
\right]  \notag \\
&&\times \mathrm{Tr}\left[ S_{c}^{bb^{\prime }}(x)\widetilde{S}%
_{c}^{aa^{\prime }}(x)+S_{c}^{ab^{\prime }}(x)\widetilde{S}_{c}^{ba^{\prime
}}(x)\right] .  \label{eq:OPE1}
\end{eqnarray}%
Here, $S_{c}(x)$ and $S_{u(d)}(x)$ are the heavy $c$- and light $u(d)$-quark
propagators, explicit expressions of which can be found, for example, in
Ref.\ \cite{Sundu:2018uyi}. In Eq.\ (\ref{eq:OPE1}), we also introduce the
shorthand notation
\begin{equation}
\widetilde{S}(x)=CS^{T}(x)C.  \label{eq:Notation}
\end{equation}

By equating the amplitudes $\Pi ^{\mathrm{Phys}}(p^{2})$ and $\Pi ^{\mathrm{%
OPE}}(p^{2})$, applying the Borel transformation to both sides of this
expression, and performing the continuum subtraction, we get an equality,
which can be used to derive sum rules for the mass $m_{T}$ and coupling $%
f_{T}$. The Borel transformation suppresses the contribution of higher
resonances and continuum states and generates a dependence of the sum rules
on a new parameter $M^{2}$. The continuum subtraction allows one, by
invoking the assumption on the quark-hadron duality, to replace an unknown
physical spectral density $\rho ^{\mathrm{Phys}}(s)$ by $\rho ^{\mathrm{OPE}%
}(s)$, which is calculable as an imaginary part of $\Pi ^{\mathrm{OPE}}(p)$.
A price paid for this simplification is appearance in the sum rules the
continuum threshold parameter $s_{0}$ that separates from one another the
ground-state and continuum contributions to $\Pi ^{\mathrm{OPE}}(p^{2})$.

To derive the final sum rules, we use this equality as well as one obtained
from the first expression by applying the operator $d/d(-1/M^{2})$. As a
result, we get
\begin{equation}
m_{T}^{2}=\frac{\int_{4m_{c}^{2}}^{s_{0}}dss\rho ^{\mathrm{OPE}%
}(s)e^{-s/M^{2}}}{\int_{4m_{c}^{2}}^{s_{0}}ds\rho ^{\mathrm{OPE}%
}(s)e^{-s/M^{2}}},  \label{eq:Mass}
\end{equation}%
and%
\begin{equation}
f_{T}^{2}=\frac{4m_{c}^{2}}{m_{T}^{4}}\int_{4m_{c}^{2}}^{s_{0}}ds\rho ^{%
\mathrm{OPE}}(s)e^{(m_{T}^{2}-s)/M^{2}}.  \label{eq:Coupl}
\end{equation}

As we have noted above, Eqs.\ (\ref{eq:Mass}) and (\ref{eq:Coupl}) depend
the auxiliary parameters $M^{2}$ and $s_{0}$. Their values are related to a
problem under analysis and should be fixed to satisfy constraints, which we
explain below. But the sum rules contain also various vacuum condensates
that are universal for all problems:%
\begin{eqnarray}
&&\langle \bar{q}q\rangle =-(0.24\pm 0.01)^{3}\ \mathrm{GeV}^{3},\   \notag
\\
&&m_{0}^{2}=(0.8\pm 0.1)\ \mathrm{GeV}^{2},\ \langle \overline{q}g_{s}\sigma
Gq\rangle =m_{0}^{2}\langle \overline{q}q\rangle ,  \notag \\
&&\langle \frac{\alpha _{s}G^{2}}{\pi }\rangle =(0.012\pm 0.004)\,\mathrm{GeV%
}^{4},  \notag \\
&&\langle g_{s}^{3}G^{3}\rangle =(0.57\pm 0.29)~\mathrm{GeV}^{6}.
\label{eq:Parameters}
\end{eqnarray}%
In numerical computations we use this information on vacuum condensates and
the $c$-quark mass $m_{c}=1.275_{-0.035}^{+0.025}\ \mathrm{GeV}$. Our
studies prove that the working regions for the parameters
\begin{equation}
M^{2}\in \lbrack 4,\ 6]~\mathrm{GeV}^{2},\ s_{0}\in \lbrack 20,\ 22]~\mathrm{%
GeV}^{2}  \label{eq:Wind1}
\end{equation}%
meet all restrictions imposed on $M^{2}$ and $s_{0}$.

The regions (\ref{eq:Wind1}) are extracted from analysis of a pole
contribution to the correlator and convergence of the sum rules. The pole
contribution (\textrm{PC})
\begin{equation}
\mathrm{PC}=\frac{\Pi (M^{2},s_{0})}{\Pi (M^{2},\infty )},  \label{eq:PC}
\end{equation}%
where $\Pi (M^{2},s_{0})$ is the Borel-transformed and subtracted invariant
amplitude $\Pi ^{\mathrm{OPE}}(p^{2})$, is one of the important quantities
necessary to extract limits of the Borel parameter $(M_{\mathrm{min}}^{2},M_{%
\mathrm{max}}^{2})$. In accordance with our computations at $M_{\mathrm{min}%
}^{2}=4~\mathrm{GeV}^{2}$, the pole contribution amounts to $0.7$, whereas
at $M_{\mathrm{max}}^{2}=6~\mathrm{GeV}^{2}$ it is $0.37$. But at the same
time, a lower limit of the Borel parameter depends on the convergence of the
operator product expansion (OPE). Restrictions imposed on $M^{2}$ by
convergence of OPE can be analyzed by means of the ratio%
\begin{equation}
R(M_{\mathrm{min}}^{2})=\frac{\Pi ^{\mathrm{DimN}}(M_{\mathrm{min}%
}^{2},s_{0})}{\Pi (M_{\mathrm{min}}^{2},s_{0})}.  \label{eq:Converg}
\end{equation}%
Here $\Pi ^{\mathrm{DimN}}(M^{2},\ s_{0})$ is a contribution to the
correlation function arising from the last term (or from the sum of last few
terms) in OPE. Numerical analysis proves that for $\mathrm{DimN}=\mathrm{%
Dim(8+9+10)}$ this ratio is $R(4~\mathrm{GeV}^{2})=0.02$, which guarantees
the convergence of the sum rules. It is worth noting that the lower boundary
of the Borel window is determined from joint analysis of \textrm{PC} and $%
R(M_{\mathrm{min}}^{2})$, i.e., the maximum accessible pole contribution is
limited by the convergence of the OPE. Additionally, at the minimum of the
Borel parameter the perturbative term amounts to $68\%$ of the total result
and exceeds the nonperturbative contributions.

In general, quantities extracted from the sum rules should not depend on the
auxiliary parameters $M^{2}$ and $s_{0}$. In real calculations, however, we
observe a residual dependence of $m_{T}$ and $f_{T}$ on them. Hence, the
choice of $M^{2}$ and $s_{0}$ should minimize these nonphysical effects as
well. The working windows for the parameters $M^{2}$ and $s_{0}$ also
satisfy these conditions. In Figs.\ \ref{fig:MassT} and \ref{fig:CouplT} we
plot the mass $m_{T}$ and coupling $f_{T}$ as functions of $M^{2}$ and $%
s_{0} $, which allows one to see uncertainties generated by the sum rule
computations. It is seen that both $m_{T}$ and $f_{T}$ depend on $M^{2}$ and
$s_{0}$, which are main sources of the theoretical uncertainties inherent in
the sum rule computations. For the mass $m_{T}$, these uncertainties are
small, $\pm 4\%$, because the ratio in Eq.\ (\ref{eq:Mass}) cancels some of
these effects. But even for the coupling $f_{T}$, the ambiguities do not
exceed $\pm 20\%$ of the central value.
\begin{widetext}

\begin{figure}[h!]
\begin{center}
\includegraphics[totalheight=6cm,width=8cm]{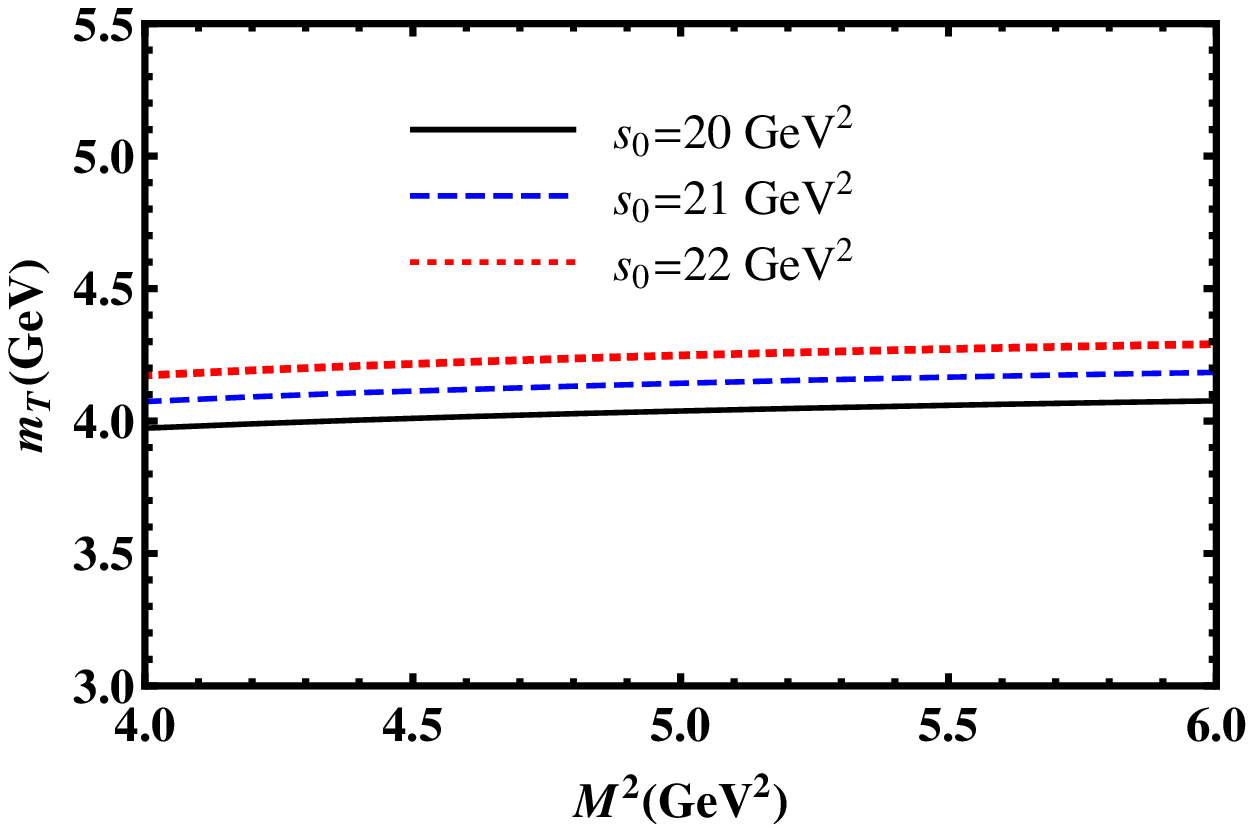}
\includegraphics[totalheight=6cm,width=8cm]{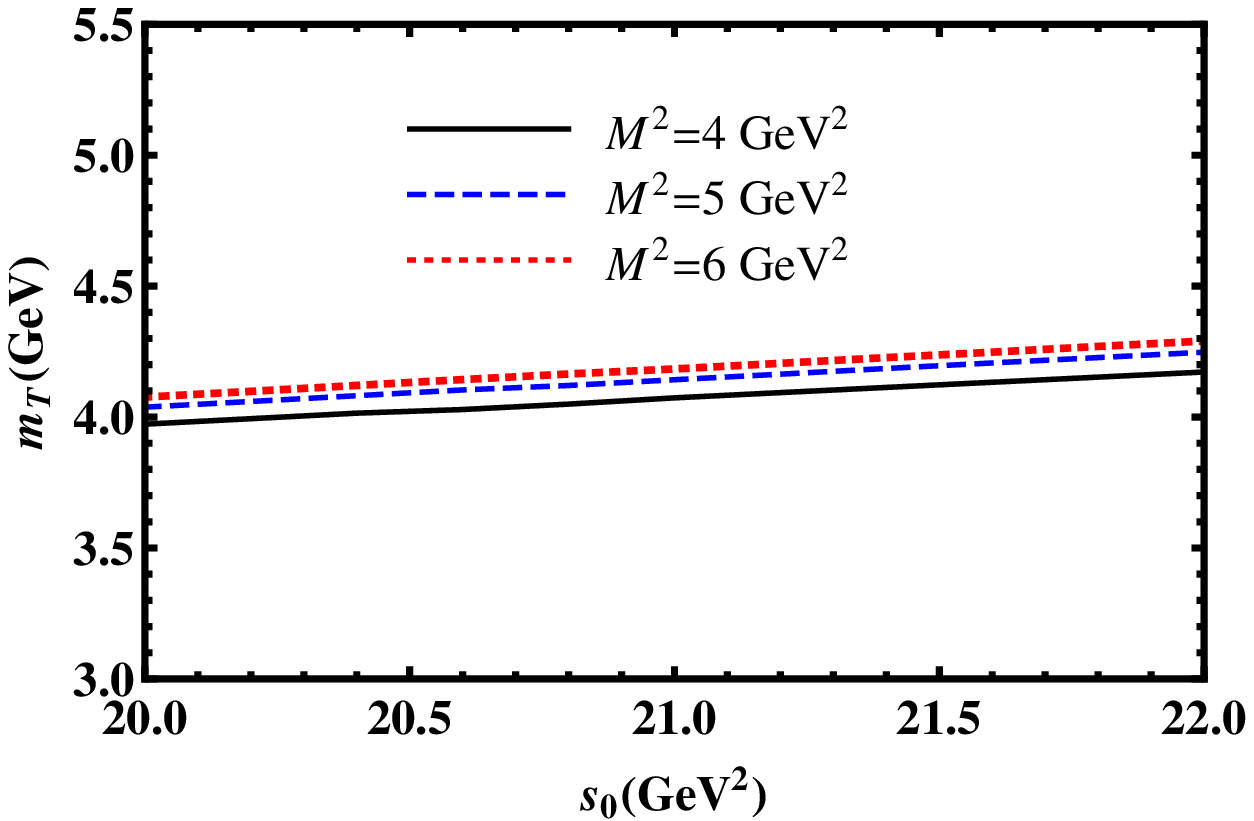}
\end{center}
\caption{ The mass of the tetraquark $T$ as a function of the Borel parameter
$M^2$ (left panel) and as a function of the continuum threshold
$s_0$ (right panel).}
\label{fig:MassT}
\end{figure}
\begin{figure}[h!]
\begin{center}
\includegraphics[totalheight=6cm,width=8cm]{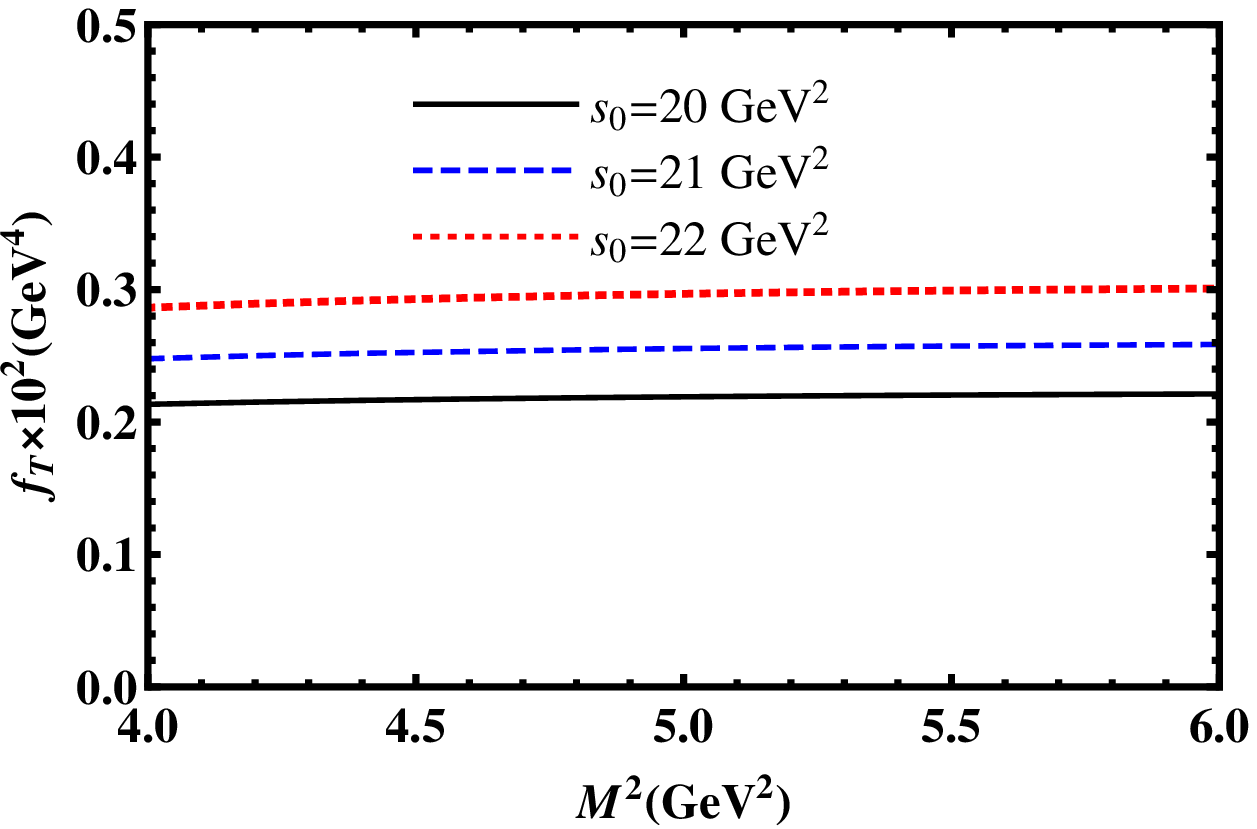}
\includegraphics[totalheight=6cm,width=8cm]{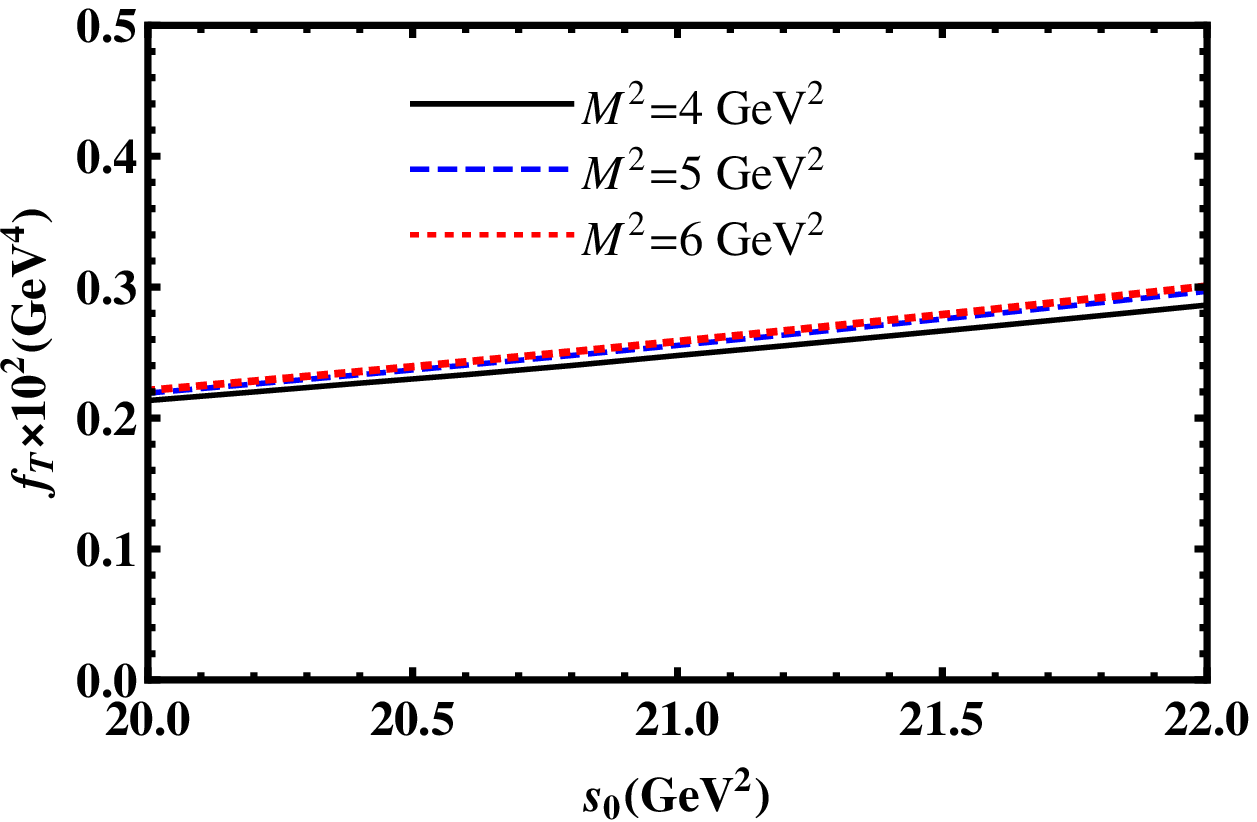}
\end{center}
\caption{ The same as in Fig.\ 1, but for the coupling $f_{T}$ of the state $T$.}
\label{fig:CouplT}
\end{figure}

\end{widetext}

Our calculations lead to the following results:
\begin{eqnarray}
m_{T} &=&(4130~\pm 170)~\mathrm{MeV},  \notag \\
f_{T} &=&(0.26\pm 0.05)\times 10^{-2}\ \mathrm{GeV}^{4}.  \label{eq:Results1}
\end{eqnarray}%
The prediction for $m_{T}$ confirms that $T$ can be interpreted as a member
of the multiplet formed by the double-charmed pseudoscalar tetraquarks. In
fact, parameters of other members of this multiplet $T_{cc;\overline{s}%
\overline{s}}^{++}$ and $T_{cc;\overline{d}\overline{s}}^{++}$ were
calculated in Ref. \cite{Agaev:2018vag}. The mass splitting between these
two states $125~\mathrm{MeV}$ is caused by the replacement $\overline{s}%
\leftrightarrow \overline{d}$ in their quark contents. By similar
substitution $\overline{s}\rightarrow \overline{u}$ in $T_{cc;\overline{d}%
\overline{s}}^{++}$, one can create the tetraquark $T$ . Comparing now the
mass $4265~\mathrm{MeV}$ of $T_{cc;\overline{d}\overline{s}}^{++}$ with $%
m_{T}=4130~~\mathrm{MeV}$, we find the mass difference $135~\mathrm{MeV}$
between these two particles. In other words, the state $T$ occupies an
appropriate place in the multiplet of the double-charmed pseudoscalar
tetraquarks, which we consider an important consistency check of the present
result.

Let us also note that $m_{T}$ is considerably lower than $(4430\pm 130)~%
\mathrm{MeV}$ predicted in Ref.\ \cite{Du:2012wp} for the pseudoscalar
tetraquark with the same quark content and structure. This discrepancy
presumably stems from the quark propagators, in which some of
higher-dimensional nonperturbative terms were neglected, and also from a
choice of the working regions for the parameters $M^{2}$ and $s_{0}$.

The mass and coupling of the state $\widetilde{T}$ can be calculated by a
similar manner. The difference here is connected with the matrix element of
the scalar particle%
\begin{equation}
\langle 0|\widetilde{J}|\widetilde{T}(p)\rangle =f_{\widetilde{T}}m_{%
\widetilde{T}},  \label{eq:ME2}
\end{equation}%
which leads to the substitution $4m_{c}^{2}/m_{T}^{4}\rightarrow 1/m_{%
\widetilde{T}}^{2}$ in the sum rule for the coupling $f_{\widetilde{T}}$ (%
\ref{eq:Coupl}). The QCD side of new sum rules is given by the expression
\begin{eqnarray}
&&\widetilde{\Pi }^{\mathrm{OPE}}(p)=i\int d^{4}xe^{ipx}\epsilon \widetilde{%
\epsilon }\epsilon ^{\prime }\widetilde{\epsilon }^{\prime }\mathrm{Tr}\left[
\gamma _{\mu }\widetilde{S}_{d}^{e^{\prime }e}(-x)\right.  \notag \\
&&\left. \times \gamma _{\nu }S_{u}^{d^{\prime }d}(-x)\right] \times \left\{
\mathrm{Tr}\left[ \gamma ^{\nu }\widetilde{S}_{c}^{bb^{\prime }}(x)\gamma
^{\mu }S_{c}^{cc^{\prime }}(x)\right] \right.  \notag \\
&&\left. -\mathrm{Tr}\left[ \gamma ^{\nu }\widetilde{S}_{c}^{cb^{\prime
}}(x)\gamma ^{\mu }S_{c}^{bc^{\prime }}(x)\right] \right\} .  \label{eq:OPE2}
\end{eqnarray}%
The new function $\widetilde{\Pi }^{\mathrm{OPE}}(p)$ also modifies the
spectral density $\rho ^{\mathrm{OPE}}(s)$. The remaining steps have been
explained above, therefore, we provide final information about the range of
the parameters used in computations
\begin{equation}
M^{2}\in \lbrack 3,\ 4]\ \mathrm{GeV}^{2},\ s_{0}\in \lbrack 19,\ 21]\
\mathrm{GeV}^{2}  \label{eq:Wind2}
\end{equation}%
and obtained predictions%
\begin{eqnarray}
m_{\widetilde{T}} &=&(3845~\pm 175)~\mathrm{MeV},  \notag \\
f_{\widetilde{T}} &=&(1.16\pm 0.26)\times 10^{-2}\ \mathrm{GeV}^{4}.
\label{eq:Results2}
\end{eqnarray}%
It is necessary to note that at $M_{\mathrm{max}}^{2}=4~\mathrm{GeV}^{2}$
the pole contribution exceeds $0.16$ which is acceptable when considering
the four-quark mesons, whereas at minimum $M_{\mathrm{min}}^{2}=3~\mathrm{GeV%
}^{2}$ it reaches $0.7$. The convergence of the operator product expansion
at $M_{\mathrm{min}}^{2}=3~\mathrm{GeV}^{2}$ is also guaranteed because $R(3~%
\mathrm{GeV}^{2})=0.03$. Our result for $m_{\widetilde{T}}$ \ is very close
to the prediction $(3870~\pm 90)~\mathrm{MeV}$ obtained in Ref. \cite%
{Wang:2017dtg}.


\section{Strong decays of the tetraquarks $T_{cc;\overline{u}\overline{d}%
}^{+}$ and $\widetilde{T}_{cc;\overline{u}\overline{d}}^{+}$}

\label{sec:Decays}
Masses of the tetraquarks $T$ and $\widetilde{T}$ are large enough to make
their strong decays to ordinary mesons kinematically allowed processes. The
mass of $T$ is $(58\pm 29)~\mathrm{MeV}$ below (we refer only to central
value of $m_{T}$ ) the $S$-wave $D^{+}D_{0}^{\ast }(2400)^{0}$ threshold but
\ is $255~\mathrm{MeV}$ above the open-charm $D^{+}D^{\ast }(2007)^{0}$ and $%
D^{0}D^{\ast }(2010)^{+}$ thresholds, and, hence, $T$ can decay in $P$-wave
to these conventional mesons. The exotic state $\widetilde{T}$ decays in $S$%
-wave to a pair of $D^{+}D^{0}$ mesons because its mass $m_{\widetilde{T}}$
exceeds $110~\mathrm{MeV}$ the corresponding border. The $P$-wave decays of $%
\widetilde{T}$ require a master particle to be considerably heavier than $%
3845~~\mathrm{MeV}$, which is not the case.

Below we consider in a detailed form the decay $T\rightarrow D^{+}D^{\ast
}(2007)^{0}$ and present final results for the remaining modes. Our goal
here is to calculate the strong coupling corresponding to the vertex $%
TD^{+}D^{\ast }(2007)^{0}$. To derive the QCD three-point sum rule for this
coupling and extract its numerical value, one begins from analysis of the
correlation function
\begin{eqnarray}
\Pi _{\mu }(p,p^{\prime }) &=&i^{2}\int d^{4}xd^{4}ye^{i(p^{\prime
}y-px)}\langle 0|\mathcal{T}\{J_{\mu }^{D^{\ast }}(y)  \notag \\
&&\times J^{D}(0)J^{\dagger }(x)\}|0\rangle .  \label{eq:CF2}
\end{eqnarray}%
Here $J(x)$,$\ J^{D}(x)$ and $J_{\mu }^{D^{\ast }}(x)$ are the interpolating
currents for the tetraquark $T$ and mesons $D^{+}$ and $D^{\ast }(2007)^{0}$%
, respectively. The $J(x)$ is given by Eq.\ (\ref{eq:Curr1}), whereas for the
remaining two currents, we use
\begin{equation}
J_{\mu }^{D^{\ast }}(x)=\overline{u}^{i}(x)i\gamma _{\mu }c^{i}(x),\ \
J^{D}(x)=\overline{d}^{j}(x)i\gamma _{5}c^{j}(x).  \label{eq:Curr3}
\end{equation}%
The 4-momenta of the tetraquark $T$ and meson $D^{\ast }(2007)^{0}$ are $p$
and $p^{\prime }$; then, the momentum of the meson $D^{+}$ is $q=p-p^{\prime
}$.

We follow the standard prescriptions of the sum rule method and calculate
the correlation function $\Pi _{\mu }(p,p^{\prime })$ using both physical
parameters of the particles involved into a process and quark-gluon degrees
of freedom. Separating the ground-state contribution to the correlation
function (\ref{eq:CF2}) from contributions of higher resonances and
continuum states,\ for the physical side of the sum rule $\Pi _{\mu }^{%
\mathrm{Phys}}(p,p^{\prime })$, we get%
\begin{eqnarray}
&&\Pi _{\mu }^{\mathrm{Phys}}(p,p^{\prime })=\frac{\langle 0|J_{\mu
}^{D^{\ast }}|D^{\ast 0}(p^{\prime })\rangle \langle 0|J^{D}|D^{+}(q)\rangle
}{(p^{\prime 2}-m_{1D^{\ast }}^{2})(q^{2}-m_{D}^{2})}  \notag \\
&&\times \frac{\langle D^{+}(q)D^{\ast 0}(p^{\prime })|T(p)\rangle \langle
T(p)|J^{\dagger }|0\rangle }{(p^{2}-m_{T}^{2})}+\ldots  \label{eq:CF3}
\end{eqnarray}

The function $\Pi _{\mu }^{\mathrm{Phys}}(p,p^{\prime })$ can be further
simplified by expressing matrix elements in terms of the mesons' physical
parameters. To this end, we introduce the matrix elements
\begin{eqnarray}
\langle 0|J^{D}|D^{+}\rangle &=&\frac{m_{D}^{2}f_{D}}{m_{c}},\   \notag \\
\langle 0|J_{\mu }^{D^{\ast }}|D^{\ast 0}\rangle &=&m_{1D^{\ast }}f_{D^{\ast
}}\varepsilon _{\mu },  \label{eq:Mel2}
\end{eqnarray}%
where $m_{D}$,$\ m_{1D^{\ast }}$ and $f_{D}$, $f_{D^{\ast }}$ are the masses
and decay constants of the mesons $D^{+}$ and $D^{\ast }(2007)^{0}$,
respectively. In Eq.\ (\ref{eq:Mel2}) $\varepsilon _{\mu }$ is the
polarization vector of the meson $D^{\ast }(2007)^{0}$. We model $\langle
D(q)D^{\ast 0}(p^{\prime })|T(p)\rangle $ in the form%
\begin{equation}
\langle D^{+}(q)D^{\ast 0}(p^{\prime })|T(p)\rangle =g_{1}(q^{2})q_{\mu
}\varepsilon ^{\ast \mu }  \label{eq:Ver1}
\end{equation}%
and denote by $g_{1}(q^{2})$ the strong coupling of the vertex $%
T(p)D(q)D^{\ast 0}(p^{\prime }).$ Then, it is not difficult to see that
\begin{eqnarray}
&&\Pi _{\mu }^{\mathrm{Phys}}(p,p^{\prime })=g_{1}(q^{2})\frac{%
m_{D}^{2}f_{D}m_{1D^{\ast }}f_{D^{\ast }}f_{T}m_{T}^{2}}{2m_{c}^{2}(p^{%
\prime 2}-m_{1D^{\ast }}^{2})(q^{2}-m_{D}^{2})}  \notag \\
&&\times \frac{1}{(p^{2}-m_{T}^{2})}\left( \frac{m_{T}^{2}-m_{1D^{\ast
}}^{2}-q^{2}}{2m_{1D^{\ast }}^{2}}p_{\mu }^{\prime }-q_{\mu }\right) +\ldots
\label{eq:Phys2}
\end{eqnarray}%
The correlation function $\Pi _{\mu }^{\mathrm{Phys}}(p,p^{\prime })$ has
two Lorentz structures proportional to $p_{\mu }^{\prime }$ and $q_{\mu }$.
We choose to work with the invariant amplitude $\Pi ^{\mathrm{Phys}%
}(p^{2},p^{\prime 2},q^{2})$ corresponding to the structure proportional to $%
p_{\mu }^{\prime }$. The double Borel transformation of this amplitude over
variables $p^{2}$ and $p^{\prime 2}$ forms the phenomenological side of the
sum rule. To find the QCD side of the three-point sum rule, we compute $\Pi
_{\mu }(p,p^{\prime })$ in terms of the quark propagators and get
\begin{eqnarray}
&&\Pi _{\mu }^{\mathrm{OPE}}(p,p^{\prime })=i^{2}\int
d^{4}xd^{4}ye^{i(p^{\prime }y-px)}\left\{ \mathrm{Tr}\left[ \gamma _{\mu
}S_{c}^{ja}(y-x)\right. \right.  \notag \\
&&\left. \times \widetilde{S}_{c}^{ib}(-x)\gamma _{5}\widetilde{S}%
_{d}^{bi}(x)\gamma _{5}S_{u}^{aj}(x-y)\right] +\mathrm{Tr}\left[ \gamma
_{\mu }S_{c}^{jb}(y-x)\right.  \notag \\
&&\left. \left. \times \widetilde{S}_{c}^{ia}(-x)\gamma _{5}\widetilde{S}%
_{d}^{bi}(x)\gamma _{5}S_{u}^{aj}(x-y)\right] \right\} .  \label{eq:CF4}
\end{eqnarray}

The correlation function $\Pi _{\mu }^{\mathrm{OPE}}(p,p^{\prime })$ is
calculated with dimension-5 accuracy, and has the same Lorentz structures as
$\Pi _{\mu }^{\mathrm{Phys}}(p,p^{\prime })$. The double Borel
transformation $\mathcal{B}\Pi ^{\mathrm{OPE}}(p^{2},p^{\prime 2},q^{2})$,
where $\Pi ^{\mathrm{OPE}}(p^{2},p^{\prime 2},q^{2})$ is the invariant
amplitude that corresponds to the term proportional to $p_{\mu }^{\prime }$%
, constitutes the second part of the sum rule. By equating $\mathcal{B}\Pi ^{%
\mathrm{OPE}}(p^{2},p^{\prime 2},q^{2})$ and Borel transformation of $\Pi ^{%
\mathrm{Phys}}(p^{2},p^{\prime 2},q^{2})$, and performing continuum
subtraction we find the sum rule for the coupling $g_{1}(q^{2})$.

The Borel transformed and subtracted amplitude $\Pi ^{\mathrm{OPE}%
}(p^{2},p^{\prime 2},q^{2})$ can be expressed in terms of the spectral
density $\widetilde{\rho }(s,s^{\prime },q^{2})$ which is proportional to
the imaginary part of $\Pi ^{\mathrm{OPE}}(p,p^{\prime })$,
\begin{eqnarray}
&&\Pi (\mathbf{M}^{2},\mathbf{\ s}_{0},~q^{2})=\int_{4m_{c}^{2}}^{s_{0}}ds%
\int_{m_{c}^{2}}^{s_{0}^{\prime }}ds^{\prime }\widetilde{\rho }(s,s^{\prime
},q^{2})  \notag \\
&&\times e^{-s/M_{1}^{2}}e^{-s^{\prime }/M_{2}^{2}},  \label{eq:SCoupl}
\end{eqnarray}%
where $\mathbf{M}^{2}=(M_{1}^{2},\ M_{2}^{2})$ and $\mathbf{s}_{0}=(s_{0},\
s_{0}^{\prime })$ are the Borel and continuum threshold parameters,
respectively. Then, the sum rule for $g_{1}(q^{2})$ is determined by the
expression
\begin{eqnarray}
&&g_{1}(q^{2})=\frac{4m_{c}^{2}m_{1D^{\ast }}}{f_{D}m_{D}^{2}f_{D^{\ast
}}f_{T}m_{T}^{2}}\frac{q^{2}-m_{D}^{2}}{m_{T}^{2}-m_{1D^{\ast }}^{2}-q^{2}}
\notag \\
&&\times e^{m_{T}^{2}/M_{1}^{2}}e^{m_{1D^{\ast }}^{2}/M_{2}^{2}}\Pi (\mathbf{%
M}^{2},\mathbf{\ s}_{0},~q^{2}).  \label{eq:SRCoup}
\end{eqnarray}%
The coupling $g_{1}(q^{2})$ is a function of $q^{2}$ and, at the same time,
depends on the Borel and continuum threshold parameters which, for simplicity, are not shown 
in Eq.\  \ref{eq:SRCoup} as arguments of $g_{1}$. Afterwards, we introduce new variable $%
Q^{2}=-q^{2}$ and denote the obtained function as $g_{1}(Q^{2})$.

The sum rule (\ref{eq:SRCoup}) contains masses and decay constants of the
final mesons: these parameters are collected in Table \ref{tab:Param}. For
the masses of $D$ mesons we use information from Ref.\ \cite%
{Tanabashi:2018oca}. A choice for the decay constants of the pseudoscalar
and vector $D$ mesons is a more complicated task. They were calculated using
various models and methods in Refs. \cite%
{Ebert:2006hj,Bazavov:2011aa,Gelhausen:2013wia,Wang:2015mxa,Dhiman:2017urn}.
Predictions obtained in these papers sometimes differ from each other
considerably. Therefore, for the decay constant of the pseudoscalar $D$
mesons, we use the available experimental result, whereas for the vector
mesons, we use the QCD sum rule prediction from Ref.\ \cite{Wang:2015mxa}.

To carry out numerical analysis of $g_{1}(Q^{2})$, apart from the
spectroscopic parameters of $D$ mesons, one also needs to fix $\mathbf{M}^{2}
$ and $\mathbf{s}_{0}$. The restrictions imposed on these auxiliary
parameters are standard for sum rule computations and have been discussed
above. The windows for $M_{1}^{2}$ and $s_{0}$ correspond to the $T$
channels, and coincide with the working regions $M_{1}^{2}\in \lbrack 4,\
6]\ \mathrm{GeV}^{2}$ and $\ s_{0}\in \lbrack 20,\ 22]\ \mathrm{GeV}^{2}$
determined in the mass calculations. The next pair of parameters $%
(M_{2}^{2},\ s_{0}^{\prime })$ is chosen within the limits
\begin{equation}
M_{2}^{2}\in \lbrack 3,\ 5]\ \mathrm{GeV}^{2},\ s_{0}^{\prime }\in \lbrack
6,\ 8]\ \mathrm{GeV}^{2}.  \label{eq:Wind3}
\end{equation}%
The extracted strong coupling $g_{1}(Q^{2})$ depends on $\mathbf{M}^{2}$ and
$\mathbf{s}_{0}$; the working intervals for these parameters are chosen in
such a way as to minimize these uncertainties. For an example, in Fig.\ \ref%
{fig:g1}, we plot the coupling $g_{1}(Q^{2})$ as a function of the Borel
parameters $M_{1}^{2}$ and $M_{2}^{2}$. It is seen that the changing of $%
\mathbf{M}^{2}$ leads to varying of the coupling $g_{1}(Q^{2})$, which
nevertheless remains within allowed limits.

\begin{figure}[h]
\includegraphics[width=8.5cm]{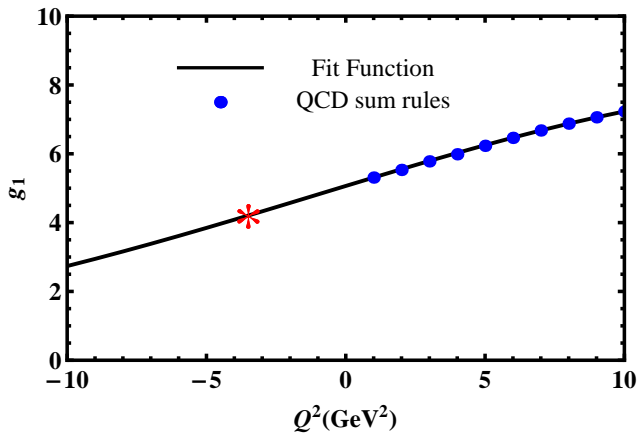}
\caption{The sum rule predictions and fit function for the strong coupling $%
g_{1}(Q^{2})$. The star shows the point $Q^{2}=-m_{D}^{2}$. }
\label{fig:Fit}
\end{figure}

The width of the decay under analysis should be computed using the strong
coupling at the $D^{+}$ meson's mass shell $q^{2}=m_{D}^{2}$, which is not
accessible to the sum rule calculations. We evade this difficulty by
employing a fit function $F_{1}(Q^{2})$ that for the momenta $Q^{2}>0$
coincides with QCD sum rule's predictions, but can be extrapolated to the
region of $Q^{2}<0$ to find $g_{1}(-m_{D}^{2})$. In the present work, to
construct the fit function $F_{1}(Q^{2})$, we use the analytic form
\begin{equation}
F_{i}(Q^{2})=F_{0}^{i}\mathrm{\exp }\left[ c_{1}^{i}\frac{Q^{2}}{m_{T}^{2}}%
+c_{2}^{i}\left( \frac{Q^{2}}{m_{T}^{2}}\right) ^{2}\right] ,
\label{eq:FitF}
\end{equation}%
where $F_{0}^{i}$, $c_{1}^{i}$ and $c_{2}^{i}$ are fitting parameters.
Numerical analysis allows us to fix $F_{0}^{1}=5.06$, $c_{1}^{1}=0.83$ and $%
c_{2}^{1}=-0.38$. In Fig.\ \ref{fig:Fit} we depict the sum rule predictions
for $g_{1}(Q^{2})$ and also provide $F_{1}(Q^{2})$; a nice agreement
between them is evident.

This function at the mass shell $Q^{2}=-m_{D}^{2}$ gives%
\begin{equation}
g_{1}\equiv F_{1}(-m_{D}^{2})=4.21\pm 0.65.  \label{eq:Coupl1}
\end{equation}%
The width of decay $T\rightarrow D^{+}D^{\ast }(2007)^{0}$ is determined by
the simple formula
\begin{equation}
\Gamma \left[ T\rightarrow D^{+}D^{\ast }(2007)^{0}\right] =\frac{%
g_{1}^{2}\lambda ^{3}\left( m_{T},m_{1D^{\ast }},m_{D}\right) }{8\pi
m_{1D^{\ast }}^{2}},  \label{eq:DW1a}
\end{equation}%
where%
\begin{equation}
\lambda \left( a,b,c\right) =\frac{1}{2a}\sqrt{a^{4}+b^{4}+c^{4}-2\left(
a^{2}b^{2}+a^{2}c^{2}+b^{2}c^{2}\right) }.
\end{equation}%
Using the strong coupling from Eq.\ (\ref{eq:Coupl1}), it is not difficult
to evaluate width of the decay$T\rightarrow D^{+}D^{\ast }(2007)^{0}$%
\begin{equation}
\Gamma \left[ T\rightarrow D^{+}D^{\ast }(2007)^{0}\right] =(64.3\pm 16.5)\
\mathrm{MeV}\text{.}  \label{eq:DW1Numeric}
\end{equation}

The second process $T\rightarrow D^{0}D^{\ast }(2010)^{+}$ can be considered
via the same manner. Corrections which should to be made in the physical
side and matrix elements of the previous decay channel are trivial. \ Thus,
the QCD side of the new sum rule in the approximation $m_{u}=m_{d}=0$
adopted in this paper coincides with $\Pi _{\mu }^{\mathrm{OPE}}(p,p^{\prime
})$. The Borel and threshold parameters $\mathbf{M}^{2}$ and $\mathbf{s}_{0}$
are chosen as in the first process. The differences are connected with the
spectroscopic parameters of produced mesons $D^{0}$ and $D^{\ast }(2010)^{+}$%
. These factors modify numerical predictions for $g_{2}(Q^{2})$, which is
the strong coupling of the vertex $TD^{0}D^{\ast }(2010)^{+}$, and change
the fit function $F_{2}(Q^{2})$. For parameters of $F_{2}(Q^{2})$, we get $%
F_{0}^{2}=5.11$, $c_{1}^{2}=0.83$, and $c_{2}^{2}=-0.38$. The result for the
partial width of the decay $T\rightarrow D^{0}D^{\ast }(2010)^{+}$ reads%
\begin{equation}
\Gamma \left[ T\rightarrow D^{0}D^{\ast }(2010)^{+}\right] =(65.6\pm 16.8)\
\mathrm{MeV}\text{.}  \label{eq:DW2numeric}
\end{equation}

The decay of the scalar four-quark meson $\widetilde{T}\rightarrow
D^{+}D^{0} $ is the last process to be considered in this section. To
extract the sum rule for the strong coupling $G(q^{2})$ corresponding to the
vertex $\widetilde{T}D^{+}D^{0}$ we start from the correlation function,
\begin{eqnarray}
&&\widetilde{\Pi }(p,p^{\prime })=i^{2}\int d^{4}xd^{4}ye^{i(p^{\prime
}y-px)}\langle 0|\mathcal{T}\{J^{D}(y)  \notag \\
&&\times J^{D^{0}}(0)\widetilde{J}^{\dagger }(x)\}|0\rangle ,  \label{eq:CF5}
\end{eqnarray}%
where $\widetilde{J}(x)$ and $J^{D}(x)$ are the interpolating currents of
the particles $\widetilde{T}$ and $D^{+}$ defined by Eqs.\ (\ref{eq:Curr2})
and (\ref{eq:Curr3}), respectively. For the interpolating current of the
pseudoscalar meson $D^{0}$, we use%
\begin{equation}
J^{D^{0}}(x)=\overline{u}^{j}(x)i\gamma _{5}c^{j}(x).  \label{eq:Curr4}
\end{equation}%
Then, it is not difficult to get the physical side of the sum rule
\begin{eqnarray}
&&\widetilde{\Pi }^{\mathrm{Phys}}(p,p^{\prime })=\frac{\langle
0|J^{D}|D^{+}(p^{\prime })\rangle \langle 0|J^{D^{0}}|D^{0}(q)\rangle }{%
(p^{\prime 2}-m_{D}^{2})(q^{2}-m_{D^{0}}^{2})}  \notag \\
&&\times \frac{\langle D^{0}(q)D^{+}(p^{\prime })|\widetilde{T}(p)\rangle
\langle \widetilde{T}(p)|\widetilde{J}^{\dagger }|0\rangle }{(p^{2}-m_{%
\widetilde{T}}^{2})}+\ldots  \label{eq:CF5A}
\end{eqnarray}%
Introducing the new matrix elements
\begin{eqnarray}
&&\langle 0|J^{D^{0}}|D^{0}(q)=\frac{m_{D^{0}}^{2}f_{D^{0}}}{m_{c}},  \notag
\\
&&\langle D^{0}(q)D^{+}(p^{\prime })|\widetilde{T}(p)\rangle
=G(q^{2})(p\cdot p^{\prime }),  \label{eq:Mel3}
\end{eqnarray}%
one can rewrite $\widetilde{\Pi }^{\mathrm{Phys}}(p,p^{\prime })$ in terms
of the physical parameters%
\begin{eqnarray}
&&\widetilde{\Pi }^{\mathrm{Phys}}(p,p^{\prime })=G(q^{2})\frac{%
m_{D}^{2}f_{D}f_{\widetilde{T}}m_{\widetilde{T}}}{2m_{c}^{2}(p^{\prime
2}-m_{D}^{2})(p^{2}-m_{\widetilde{T}}^{2})}  \notag \\
&&\times \frac{m_{D^{0}}^{2}f_{D^{0}}}{(q^{2}-m_{D^{0}}^{2})}\left( m_{%
\widetilde{T}}^{2}+m_{D}^{2}-q^{2}\right) +\ldots  \label{eq:CF5B}
\end{eqnarray}%
In Eqs.\ (\ref{eq:Mel3}) and (\ref{eq:CF5B}), $m_{D^{0}}$ and $f_{D^{0}}$
are the $D^{0}$ meson's mass and decay constant, respectively.

The QCD side of the sum rule $\widetilde{\Pi }^{\mathrm{OPE}}(p,p^{\prime })$
is given by the expression
\begin{eqnarray}
&&\widetilde{\Pi }^{\mathrm{OPE}}(p,p^{\prime })=i^{2}\int
d^{4}xd^{4}ye^{i(p^{\prime }y-px)}\epsilon \widetilde{\epsilon }\left\{
\mathrm{Tr}\left[ \gamma _{5}S_{c}^{ic}(y-x)\right. \right.   \notag \\
&&\left. \times \gamma _{\mu }\widetilde{S}_{c}^{ib}(-x)\gamma _{5}%
\widetilde{S}_{u}^{dj}(x)\gamma ^{\mu }S_{d}^{ei}(x-y)\right] -\mathrm{Tr}%
\left[ \gamma _{5}S_{c}^{ib}(y-x)\right.   \notag \\
&&\left. \left. \times \gamma _{\mu }\widetilde{S}_{c}^{jc}(-x)\gamma _{5}%
\widetilde{S}_{u}^{dj}(x)\gamma ^{\mu }S_{d}^{ei}(x-y)\right] \right\} .
\label{eq:CF6}
\end{eqnarray}%
The standard operations with $\widetilde{\Pi }^{\mathrm{Phys}}(p,p^{\prime })
$ and $\widetilde{\Pi }^{\mathrm{OPE}}(p,p^{\prime })$ yield the sum rule%
\begin{eqnarray}
&&G(q^{2})=\frac{2m_{c}^{2}}{m_{D}^{2}f_{D}f_{\widetilde{T}}m_{\widetilde{T}%
}m_{D^{0}}^{2}f_{D^{0}}}\frac{q^{2}-m_{D^{0}}^{2}}{m_{\widetilde{T}%
}^{2}+m_{D}^{2}-q^{2}}  \notag \\
&&\times e^{m_{\widetilde{T}}^{2}/M_{1}^{2}}e^{m_{D}^{2}/M_{2}^{2}}%
\widetilde{\Pi }(\mathbf{M}^{2},\mathbf{\ s}_{0},~q^{2}).
\label{eq:SRCoupl2}
\end{eqnarray}%
In numerical calculations, the auxiliary parameters for the $\widetilde{T}$
and $D^{+}$ channels are chosen as in Eqs.\ (\ref{eq:Wind2}) and (\ref%
{eq:Wind3}), respectively. The parameters of the fit function $F_{3}(Q^{2})$
are equal to $F_{0}^{3}=0.31~\mathrm{MeV}^{-1}$, $c_{1}^{3}=-1.15$, and $%
c_{2}^{3}=0.92$, which at the mass shell $Q^{2}=-m_{D^{0}}^{2}$ leads to the
strong coupling
\begin{equation}
G\left( -m_{D^{0}}^{2}\right) =(0.43\pm 0.07)\ \mathrm{GeV}^{-1}.
\end{equation}%
The width of this decay is determined by the expression
\begin{equation}
\Gamma \lbrack \widetilde{T}\rightarrow D^{+}D^{0}]=\frac{G^{2}m_{D}^{2}}{%
8\pi }\lambda \left( 1+\frac{\lambda ^{2}}{m_{D}^{2}}\right) ,
\label{eq:DW2}
\end{equation}%
where $\lambda =\left( m_{\widetilde{T}},m_{D},m_{D^{0}}\right) $. Numerical
computations predict%
\begin{equation}
\Gamma \lbrack \widetilde{T}\rightarrow D^{+}D^{0}]=(12.4\pm 3.1)\ \mathrm{%
MeV}\text{.}  \label{eq:DW2a}
\end{equation}

\begin{table}[tbp]
\begin{tabular}{|c|c|}
\hline\hline
Parameters & Values ( $\mathrm{MeV}$ ) \\ \hline\hline
$m_{D^{0}}$ & $1864.83\pm 0.05$ \\
$m_{D}$ & $1869.65\pm 0.05$ \\
$m_{1D^{*}}~(D^{*}(2007)^{0})$ & $2006.85\pm 0.05$ \\
$m_{2D^{*}}~(D^{*}(2010)^{+})$ & $2010.26\pm 0.05$ \\
$f_{D}$ & $203.7 \pm 1.1$ \\
$f_{D^{*}}$ & $263 \pm 21$ \\ \hline\hline
\end{tabular}%
\caption{Parameters of $D$ mesons produced in the decays of the tetraquarks $%
T$ and $\widetilde{T}$.}
\label{tab:Param}
\end{table}
\begin{figure}[h]
\includegraphics[width=8.8cm]{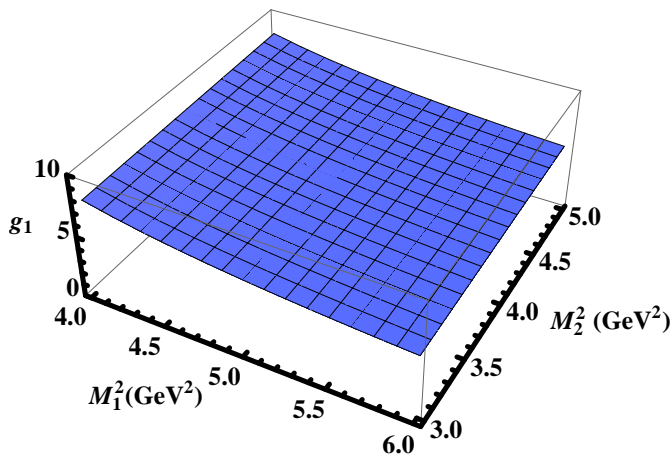}
\caption{The strong coupling $g_{1}(Q^{2})$ as a function of the Borel
parameters $\mathbf{M}^{2}=(M_{1}^{2},\ M_{2}^{2})$ at the fixed $%
(s_{0},s_{0}^{\prime })=(21,7)~\mathrm{GeV}^{2}$ and $Q^{2}=5~\mathrm{GeV}%
^{2}$.}
\label{fig:g1}
\end{figure}
The partial width of these decays are the main result of the present section.


\section{Conclusions}

\label{sec:Conc}
In this work we have explored features of the double-charmed pseudoscalar
and scalar tetraquarks $T$ and $\widetilde{T}$. We have calculated their
masses and couplings as well as found partial width of their strong decays.
Our result for $m_{T}$ has allowed us to interpret the resonance $T$ as a
member of the multiplet of double-charmed pseudoscalar tetraquarks.
Saturating the full width of $T$ by the decays $T\rightarrow D^{+}D^{\ast
}(2007)^{0}$ and $T\rightarrow D^{0}D^{\ast }(2010)^{+}$, it is possible to
find%
\begin{equation}
\Gamma _{T}=(129.9\pm 23.5)~\mathrm{MeV}.  \label{eq:FWidth}
\end{equation}%
Other members of this multiplet are tetraquarks $T_{cc;\overline{s}\overline{%
s}}^{++}$ and $T_{cc;\overline{d}\overline{s}}^{++}$, which were explored in
Ref. \cite{Agaev:2018vag}. These tetraquarks together with $T$ form\ the
correct pattern of the pseudoscalar multiplet. Indeed, masses of these
particles differ from each other by \ approximately $125~\mathrm{MeV}$,
caused by an existence or absence of $s$ quark(s) in their contents. The
full widths of the exotic mesons $\Gamma \lbrack T_{cc;\overline{s}\overline{%
s}}^{++}]=(302~\pm 113)~\mathrm{MeV}$ and $\Gamma \lbrack T_{cc;\overline{d}%
\overline{s}}^{++}]=(171~\pm 52)~\mathrm{MeV}$ are large, and we can
classify them as broad resonances. The full width of the tetraquark $T$
differs from $\Gamma \lbrack T_{cc;\overline{s}\overline{s}}^{++}]$
considerably but is comparable to $\Gamma \lbrack T_{cc;\overline{d}%
\overline{s}}^{++}]$. Therefore, we include the pseudoscalar tetraquark $T$
\ in a class of broad resonances.

The scalar double-charmed tetraquark $\widetilde{T}$ with full width $\Gamma
_{\widetilde{T}}=(12.4\pm 3.1)~\mathrm{MeV}$ is a relatively narrow state.
This resonance is a member of a double-charmed scalar tetraquarks'
multiplet. Investigation of other members of this multiplet, calculation of
their masses, and partial and full widths can provide valuable information
about properties of these scalar particles.


\section*{Acknowledgments}

S.~S.~A. is grateful to Professor V.~M.~Braun for enlightening and helpful
discussions.

\end{document}